\documentclass[sigconf]{acmart}

\usepackage{url}
\usepackage{enumitem}
\AtBeginDocument{%
  \providecommand\BibTeX{{%
    \normalfont B\kern-0.5em{\scshape i\kern-0.25em b}\kern-0.8em\TeX}}}

\newcommand{\smallsection}[1]{\textbf{#1:}\,}

\copyrightyear{2024}
\acmYear{2024}
\setcopyright{rightsretained}
\acmConference[SIGCSE 2024]{Proceedings of the 55th ACM Technical Symposium on Computer Science Education V. 1}{March 20--23, 2024}{Portland, OR, USA}
\acmBooktitle{Proceedings of the 55th ACM Technical Symposium on Computer Science Education V. 1 (SIGCSE 2024), March 20--23, 2024, Portland, OR, USA}
\acmDOI{10.1145/3626252.3630926}
\acmISBN{979-8-4007-0423-9/24/03}

\makeatletter
\gdef\@copyrightpermission{
  \begin{minipage}{0.3\columnwidth}
   \href{https://creativecommons.org/licenses/by/4.0/}{\includegraphics[width=0.90\textwidth]{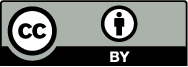}}
  \end{minipage}\hfill
  \begin{minipage}{0.7\columnwidth}
   \href{https://creativecommons.org/licenses/by/4.0/}{This work is licensed under a Creative Commons Attribution International 4.0 License.}
  \end{minipage}
  \vspace{5pt}
}
\makeatother

\begin{document}

\title{Socially Responsible Computing in an Introductory Course}


\author{Aakash Gautam}
\affiliation{%
  \institution{University of Pittsburgh}
  \country{Pittsburgh, PA, USA}
}
\email{aakash@pitt.edu}

\author{Anagha Kulkarni}
\affiliation{%
  \institution{San Francisco State University}
  \country{San Francisco, CA, USA}
}
\email{ak@sfsu.edu}

\author{Sarah Hug}
\affiliation{%
  \institution{CERC}
  \country{Westminster, CO, USA}
}
\email{hug@colorado.edu}

\author{Jane Lehr}
\affiliation{%
  \institution{Cal Poly San Luis Obispo}
  \country{San Luis Obispo, CA, USA}
}
\email{jlehr@calpoly.edu}

\author{Ilmi Yoon}
\affiliation{%
  \institution{San Francisco State University}
  \country{San Francisco, CA, USA}
}
\email{ilmi@sfsu.edu}


\renewcommand{\shortauthors}{Aakash Gautam, Anagha Kulkarni, Sarah Hug, Jane Lehr, \& Ilmi Yoon}

\begin{abstract}

Given the potential for technology to inflict harm and injustice on society, it is imperative that we cultivate a sense of social responsibility among our students as they progress through the Computer Science (CS) curriculum.
Our students need to be able to examine the social complexities in which technology development and use are situated. 
Also, aligning students' personal goals and their ability to achieve them in their field of study is important for promoting motivation and a sense of belonging. 
Promoting communal goals while learning computing can help broaden participation, particularly among groups who have been historically marginalized in computing. 
Keeping these considerations in mind, we piloted an introductory Java programming course in which activities engaging students in ethical and socially responsible considerations were integrated across modules.
Rather than adding social on top of the technical content, our curricular approach seeks to weave them together.
The data from the class suggests that the students found the inclusion of the social context in the technical assignments to be more motivating and expressed greater agency in realizing social change.
We share our approach to designing this new introductory socially responsible computing course and the students' reflections. 
We also highlight seven considerations for educators seeking to incorporate socially responsible computing.

\end{abstract}





\maketitle

\section{Introduction}


``Undergraduates need to understand the basic cultural, social, legal, and ethical issues inherent in the discipline of computing ... They should understand their individual roles in this process'' 
argue the 1991 Report of the ACM/IEEE-CS Joint Curriculum Task Force \cite{curricula1991report}. 
Similar arguments to integrate ethics have been repeatedly made over the years (e.g., \cite{werth1997getting, martin1999awareness, buckley2004benefits, califf2005effective, connolly2011beyond, goldweber2011enhancing, johnson1994should, pauca2012mobile, cohen2021new, medina2021towards, zegura2023moment, castro2023piloting}).
The growing awareness of computing systems' potential to inflict harm and injustice on society has strengthened the argument for integrating ethical responsibilities in computing curricula \cite{ko2020time, fiesler2020we, yadav2022breaking}.

Equally importantly, education researchers have long since argued for the importance of ensuring \emph{goal congruity}, which posits that students need to perceive an alignment between their personal goals and their ability to fulfill those goals by participating in the field of study (called goal affordances) for them to pursue a career path in that field \cite{diekman2010seeking}. 
In computing education, a greater emphasis on agentic goals, which has an inward focus on ``the self, self-efficacy, and working with things (instead of people)'' in contrast to communal goals, which has an outward focus on ``working with, or in service of others'' has been found to be a barrier in enhancing diversity and inclusion in computing \cite{brinkman2016applying, boucher2017can, stewart2022equity}. 
Evidence from early CS courses that promoted affordances to meet communal goal have been shown to be successful in increasing participation of women, African American, and Hispanic students \cite{brinkman2016applying, corbett2015solving, pinkard2020equitable}.

Despite repeated calls and well-established benefits, we continue to struggle in incorporating ethics in CS. 
Scholars have highlighted persistence challenges such as lack of faculty expertise \cite{fiesler2021integrating, saltz2019integrating}, faculty resistance \cite{saltz2019integrating, martin1999awareness}, time and resource constraints \cite{fiesler2021integrating, brown2022shortest}, and the perception that ethics is outside the scope of computing \cite{cohen2021new, smith2023incorporating}. 
Efforts to integrate ethics into CS courses are ongoing, with various models and solutions being proposed, ranging from a stand-alone course to incorporating it across CS courses \cite{fiesler2020we, cohen2021new, horton2023more, dean2022teaching, kiesler2023socially}.
While we have yet to fully integrate ethics in CS, we also know that ethics integration is not enough. 
We need praxis-oriented computing courses that build upon ethical considerations toward encouraging students to take responsibility by understanding the power and social impact of technology --- that is, engaging with socially responsible computing. 


\textbf{Socially responsible computing} goes beyond ethical considerations. 
It acknowledges that computing enacts a form of power, thus computing professionals need to develop a critical understanding of how technology can perpetuate or challenge societal inequities \cite{cohen2021new, ryoo2021happens, kiesler2023socially}. 
The goal is to support computing students, who are going to be builders of technology in the near future, to develop a sense of responsibility toward the systems they create.

We are embarking on a long journey to incorporate socially responsible computing across major CS courses, beginning with a new CS0 course.
The CS0 course was piloted in Spring 2023 with the plan of iterating on it and making it mandatory for all CS students in Fall 2023. 
In this paper, we share an overview of the CS0 course structure and details from the students' reflections. 
The findings highlight the value of integrating socially responsible computing to facilitate students' understanding of computing in a social context and in deepening their awareness of justice and power relations. 
We also show that students found the integration to be motivating and examined their responsibilities and structural issues surrounding computing.
Overall, the integration facilitated a synergistic opportunity to deepen the learning of both programming and social issues.
We conclude by reflecting on seven key considerations for designing a socially responsible computing course.

\vspace{-0.5em}
\section{Our Curricular Approach}



The course was structured such that students first developed a framework to examine the impact of computing and then learned to program computing systems.

\vspace{-0.5em}
\subsection{Computing Around Us}
The first three weeks were about programmed systems, not programming. This was to show students the value and power of computing, and with it develop ability to examine the impact of computing. 
We started by having students observe how their family, friends, and others used technology. 
Students had to ask people how they used technology for the assignment. 
This allowed us to discuss technology use in daily life, how designed systems affect people, and technology developers' power and responsibilities.

We then covered ethical reasoning, power, and social impact analysis.
We read and discussed Consequentialism (particularly utilitarianism), deontological ethics, and virtue ethics.
The discussions underscored the inevitability of differing beliefs and emphasized the importance of understanding differences in underlying systems of values and the various forces that influence different actors. 
Highlighting that actions are highly contextualized, we stressed the importance of considering multiple perspectives, evaluating the impact on multiple stakeholders, and carefully considering who might be ignored or harmed by technological systems. 

Students' discussions were structured by asking questions about access and power. Four questions were often presented to the students: who has access to the system and who is ignored, whose data is being collected, who is benefiting the most, and who gets to decide what problems need to be solved.
We encouraged students to consider impact on individual, communal, and societal levels.


\vspace{-0.5em}
\subsection{Computing By Us \emph{and} For Us}

After three weeks, we began teaching Java programming as in a typical CS0 course (see Table \ref{tab:course-sequence}).
We relied on three practices to integrate socially responsible computing: using socially grounded assignments, introducing  projects that intertwined social and technical issues, and encouraging individual and collective reflections.

\begin{table}[]
\resizebox{\columnwidth}{!}{%
\begin{tabular}{@{}l|l@{}}
\toprule
\textbf{Weeks} & \textbf{Major Topics}                                                                  \\ \midrule
1--3           & Stakeholders, power, ethical reasoning, and   responsibilities    \\
4--7           & Programming: Variables, sequential execution, conditional statements                   \\
8              & Bringing all together: Project 1: Evaluating On-Campus Housing Allocation              \\
9--10          & Programming: Repeating Actions through Loops                                           \\
11             & Bringing all together: Project 2: Fairly Dividing Restaurant's Pooled Tip              \\
12--15         & Programming: Arrays and Modular Programming                                            \\
16             & Bringing all together: Project 3: Filtering Interview Candidates \\ \bottomrule
\end{tabular}%
}
\caption{Major topics covered in the 16-week long course}
\vspace{-3em}
\label{tab:course-sequence}
\end{table}

\subsubsection{Socially-Grounded Assignments}

We wanted to introduce computing as a personally-meaningful and socially-relevant tool. To that end, we provided multiple opportunities for students to see computing in a familiar, socially-situated context. For example, in a lecture on loops, we worked together to create a digital diary for ourselves. 
Similarly, in a conditional statement assignment, students were required to devise a method to assist their friends in purchasing birthday gifts. This task required students to keep track of their current preferences as well as their friend's budget. 

Many tasks were centered on family and friends. 
For example, when we introduced user input processing with Java's Scanner Object, we assigned them to build a rudimentary conversational agent.
We encouraged students to make prompts in their native language to motivate them to involve family and friends. 
We provided an example in Nepali, the instructor's mother tongue, so that students could work on the topic in their native language. 
Along with English, students submitted responses in Spanish, Bahasa Indonesia, and Korean. 
Similarly, the freedom to steer the conversation in any direction appears to have encouraged students to include others, such as in inquiring about their Pokemon collections, multiple sports interests, and restaurant recommendations.


\subsubsection{Projects to Examine Social \textbf{and} Technical}
We introduced three two-week-long projects to help students connect their newly acquired technical knowledge to familiar social issues.

\smallsection{Project 1}
The first project asked students to evaluate and redesign our university's first-come, first-served on-campus housing allocation system.  
Students assessed whether the existing approach is fair, who benefits, and who is harmed.  They were asked to speak with others outside of class, finalize key factors for on-campus housing allocation, and justify those choices. They implemented a Java program codifying their proposed system, which required them to demonstrate mastery of conditional statements.

\smallsection{Project 2}
This project entailed determining an equitable division of a restaurant's pooled tip amount. Some students had mentioned in previous conversations that they worked in restaurants, so we used that as the context for the project. The assignment required students to use loops to process an indeterminate pile of restaurant bills. Students were asked to consult with others to finalize the factors they wanted to consider when dividing the pooled tip amount. After justifying their decisions on what constitutes a fair division, they implemented their algorithm for tip distribution.

\smallsection{Project 3}
Students created filters to shortlist CS job applicants for a fictitious company. Students were required to use arrays and demonstrate mastery of methods. This project provided a scaffolded exploration of some of the courses available in their pathway to graduation. We provided a hypothetical filtering system that evaluated the candidates based on their grades in CS courses. Initially, they crafted filters based solely on CS course grades. Then we asked them to talk to people both inside and outside the class to finalize a justified, fairer evaluation criteria. They then implemented the filter for shortlisting the fictional company's candidates.

The first and third projects were adapted from assignments by Evan Peck \cite{peck2017ethical}, with the second having a similar structure. 
We are happy to share all of our assignments upon request, and plan to follow Evan's footstep in making them publicly available soon.

\subsubsection{Individual and Class-wide Reflections}

We added reflective questions about power to most assignments and all projects to encourage students to think critically about the problem and their approach. We asked students to assess who benefits and who is harmed by their approach, as well as any limitations. 
In class, students reinforced these reflections by 
discussing these questions. 


For example, near the end of Project 2, we discussed various methods for dividing the pooled tip amount. Some mentioned that they talked to people who waited tables and dealt with customers and felt they should be paid more. Others, drawing from their own restaurant experiences, advocated for equal division between front and back staff, while a small group believed in equity-based distribution, with lower earners receiving more. This sparked discussions on technology's role in ensuring transparency and fairness as well as the pitfalls of technological solutionism, prompting one student to describe Project 2 as ``\textit{a good programming practice with a backdrop of a tone-deaf~\footnote{This is an ableist word that the instructor did not catch when it was shared in the survey reflection. This underscores the importance of being mindful of language practices, a crucial aspect of teaching socially responsible computing.} social issue.}''

\subsection{Data Sources and Analysis}

We present data from a pilot class taught in Spring 2023. 
The course was elective for CS students and open to non-majors.
We had 45 students, 21 of whom were computer science students and 17 were studying something other than natural sciences or engineering.
Of the 45, 12 were female and 16 were what San Francisco State University categories as under-represented minorities (URM\footnote{Our University designates URM as students whose race/ethnicity is ``African Americans, American Indians/Alaska Natives, and Latinos''.  
}).
14 students received Pell Grants and 13 were first-generation college students. 
Post each project, students filled an anonymous survey about the assignments and their perceived roles. 
The survey had 20 rating questions ranging from 0 to 100, categorized into four groups: Understanding of Computing in a Social Context, Awareness of Justice and Power Relations, Personal Relevance and Responsibilities, and Learning and Conceptual Integration. Open-ended responses were analyzed using thematic analysis \cite{saldana2015coding}.

\section{Findings}

The surveys were optional and the number of responses declined over the three surveys, with 31, 17, and 10 responses respectively (see Table \ref{tab:likert-items})\footnote{Please find the summary of each individual survey here: \url{https://osf.io/5bstn/?view_only=ba371971bcc94d44830921d3e55406fc}}.
The significant non-response rates suggests that there may be non-response bias \cite{national2013nonresponse}.
Moreover, even though the survey were anonymous, the power differences in the class setting may have influenced the response.
Noting these limitations, we do not make claims of generalizability but rather provide initial insights on the synergistic value of integrating socially responsible computing in programming classes. 

Overall, the survey responses indicate that the students recognized the social impact potential of computing, showed awareness of justice-related considerations, felt a moderate personal connection to the projects, and acknowledged the impact of socially relevant computing on their learning experience.
We now report on the students' reflections from the surveys.

\begin{table}[]
\resizebox{\columnwidth}{!}{%
\begin{tabular}{@{}llll@{}}
\toprule
\textbf{Likert  Scales}                         & \textbf{\begin{tabular}[c]{@{}l@{}}Project 1\\ (n = 31)\end{tabular}} & \textbf{\begin{tabular}[c]{@{}l@{}}Project 2\\ (n = 17)\end{tabular}} & \textbf{\begin{tabular}[c]{@{}l@{}}Project 3\\ (n = 10)\end{tabular}} \\ \midrule
Understand Computing in a Social Context & 77.3               & 82.2               & 79.4               \\
Awareness of Justice and Power Relations          & 74.7               & 68.9               & 67.9               \\
Personal Relevance and Responsibilities         & 66.1               & 62.2               & 64.3               \\ 
Learning and Conceptual Integration            & 67.4               & 74.1               & 64.5               \\
\bottomrule
\end{tabular}%
}
\caption{Mean rating scores of the Likert scales (out of 100).}
\vspace{-4em}
\label{tab:likert-items}
\end{table}

\subsection{Understand Computing in a Social Context}

Students expressed appreciation for addressing relatable ``real-world'' challenges, prompting one student to comment, ``\textit{I think the main thing I learned was how programming is used to look into real ethical problems, which I think is really cool...}'' 
They stated that the ability to see computing in a broader social context was helpful in appreciating the value of programming knowledge. 
A student aptly captured this sentiment, ``\textit{I love the idea that things go hand in hand with programming, because programming can be under a broad number of things in society. 
}''
They also valued consulting with stakeholders, recognizing it as an opportunity to understand the constraints of computing against systemic problems.

\subsubsection{Valuing working on and learning from relatable problems}
Throughout the projects, students appreciated the opportunity to work on relatable real-world problems. As one student reflected on Project 1, ``\textit{It was meaningful to solve problems directly related to reality through programming.}'' 
The importance of relatable tasks was emphasized by another student, who stated, ``\textit{I really enjoyed working on something that was somewhat relevant to us.}''

Some projects were more relatable than others. Reflecting on Project 2, a student remarked,``\textit{This was by far my favorite one of the two, just because I could relate to it so I had a good idea of where to start.}''
Similarly, on Project 3, another student noted its future relevance: ``\textit{This is going to be something I'm gonna be on the other end of when I start applying for jobs so I hope whoever wrote the algorithm was at least more forgiving than me.}''
In contrast, the absence of perceived social impact was discouraging. As one student reflected on Project 2, ``\textit{I couldn’t see the social impact of it like I could in the first project. It felt like I was making something someone wanted versus making something I felt made a difference. I felt unmotivated while working and was mostly interested in making the code work rather than why I was writing the code in the first place.}''

\subsubsection{Engaging with stakeholders to understand the impact of computing}
Engaging with external stakeholders was central to the projects. 
To help with that, early assignments encouraged students to converse with various individuals and observe their technological interactions. 
Students generally appreciated this engagement. Yet, a few encountered challenges, as highlighted by a student's reflection, ``\textit{the only challenges I faced when working on this project was finding people who wanted to be interviewed. I had to introduce myself as taking a survey for student housing.}'' 

Some displayed commendable initiative. 
One recounted spending hours on campus, questioning passersby about housing, successfully engaging with 11 individuals. 
While many conversed with friends or family, a few expanded their search to community members. 
In reflecting on this, one student summarized, ``\textit{Some students expressed they feel like it’s weird that we have to interview, but it kinda makes sense since we are trying to solve a real world issue.}''

Many shared gaining new perspectives through the interactions, as expressed by a student who grappled with the complexities of on-campus housing, ``\textit{I had a harsh reality of how expensive housing on campus is and I did not know how to support financially weaker backgrounds by giving them more incentive to live on campus since it is in fact not affordable. That was a challenge that I faced while I was working on various aspects.}''

\vspace{-1em}
\subsection{Awareness of Justice and Power Relations}

Overall, the students expressed positive value in learning about the social complexities tied to the project.  For example, in response to the on-campus housing issue, a student wrote, ``\textit{I got} [to] \textit{learn specifically housing allocation issue at SF since I was unaware being a commuter. It was interesting to learn about the process of it.}''
Through the projects, students grappled with power, especially developers' power and computing limitations in the face of structural issues.

\subsubsection{Reflecting on developers' power-to and power-over}

The social context of the projects prompted students to consider the power software developers have in making decisions. Early semester discussions, centered on feminist perspectives on power---specifically the concepts of ``power-to'' and ``power-over'' \cite{allen2022feminist}---provided a foundation for students to evaluate power dynamics critically. 

As students worked on the projects, they began to reflect on the power they exercised based on the factors they considered and the decisions they integrated into their programs. 
One student, for example, stated, ``\textit{...in this case it was a problem relevant to me because my peers are also affected by the housing situation. I learned that there are many factors within the problem, and the more factors there is then either people are drowned out or helped even more.}''
Another student, reflecting on Project 1, wrote about the power, ``\textit{The biggest takeaway from this project was understanding how the choices we make as programmers can influence the lives of others.}''

Power dynamics became even more evident to some students during Project 2. One of them remarked, "\textit{I learned how computing has real world power and can really affect real world people for good or bad reasons.}''
Drawing from previous lessons, another student delved into the manifestation of these concepts in software, expressing, ``\textit{I never really thought about the connection between power to and power over in program or how it would be excited} [executed] \textit{in program. So it was interesting to see how it works.}''

\subsubsection{Examining the limitations of computing as a technical fix to social problems}
Our discussions continuously returned to the structural challenges. 
Our aim was for students to recognize and critically examine the larger structures within which technology functions \cite{ko2020time}. 
By foregrounding these structural considerations, we hoped to challenge the pervasive notion of techno-solutionism and instead emphasize the importance of collective action.

As students worked on their projects, they began to assess the limitations of computing.
One student, during their work on Project 3, shared the reductionism they grappled with when having to represent complex human attributes within computational frameworks: ``\textit{I was challenged to think about how to quantize} [sic] \textit{factors such as a candidate's gender and whether they are a veteran. Veteran status can easily be reduced to a binary 1 or 0, but gender/sex is a much more difficult concept to describe using integers. I was forced to resort to a simplified binary view of gender/sex which is obviously counterproductive in light of modern norms around gender identity.}''
Such reflections were not isolated. Pondering the scope and constraints of computing in addressing complex societal issues, a student from Project 2 observed, ``\textit{The programming experience was valuable. However, the social problem in this case is much deeper than anything that can be addressed by computing - and is in fact exacerbated by computing in the real world. This is the issue of our relationship to labor. Systemic changes are required to address problems like worker exploitation and the subsidization of wages with tips} ...''

As the semester progressed, a noticeable trend emerged: students increasingly drew connections between structural issues and their technological development. Reflecting on the ethical quandaries of devising systems like candidate filtering algorithms, a student insightfully remarked on the larger societal inequities exacerbated by unchecked technological advancements, ``\textit{I think that technology is actually something that exacerbates the inequities of the world. The tech sector enriches a small group of individuals while the rest of the world continues to languish in poverty. Technology in a way is the inevitable outcome of Enlightenment philosophy where the entire natural world became something for humans to exercise power over. The laptop I am typing this on required large amounts of raw material mined from distant parts of the world by workers subject to horrendous work conditions. Human progress and technological progress have become conflated. See the enduring struggle of the American proletariat despite record-high, tech-enabled productivity.}''

\vspace{-0.5em}
\subsection{Personal Relevance and Responsibilities }

Similarly, students deeply considered their roles in addressing societal challenges in assignments and projects. 
They recognized their unique positions not just as coders, but as socially responsible individuals. Many found resonance in the course’s emphasis on real-world issues. One student expressed, ``\textit{... this which gave me an opportunity to apply myself and made me look at things like a human, a programmer and not a robot designed to work mechanically. It allowed for me to think for others and address different issues.}'' 

The students' reflections suggested their evolving understanding of their social responsibilities. One student, pondering on their Project 1 experience, shared, ``\textit{While working on the project, I learnt how to put my social ideas into implementation through programming and also got to understand the power that programmers hold at various platforms in today's society. It made me understand how important our ethical values are.}''
The emphasis on developer's ethical values indicates our success in fostering a reflective environment. 
Similar several reflective discussions throughout the semester revolved around choices and the decisions to balance multiple tensions.

\subsubsection{Grappling with choices and decisions}

Students engaged with the nuances of equality, fairness, and equity. A student's reflection from Project 1 captures this tension, ``\textit{I learned how to create a point system in java dependent on user answers. I also learned the weight/importance I put on my factors which can easily drown out fairness. I approached the assignment valuing equity more than fairness.}''
Many shared difficulties in iterating towards their vision of a fair outcome, ``\textit{Over time, through building the algorithm, it dawned on me that more factors likely provided more fairness.}'' 

The projects required students to navigate multiple tensions that are inherent in any complex social problem, highlighting the significant power and responsibilities accorded to computer programmers. 
Reflecting on their power and their internal conflict in making decisions, a student noted, ``\textit{It was difficult for me to decide on what I would give more points to and what I would penalize in the program as I was unsure what groups I wanted to benefit the most.}''

Indeed, achieving a fair balance was cited as one of the most challenging parts of the projects.
Reflecting on Project 2, a student wrote, ``\textit{I was struggling with how to divide the tips among the workers provided and I still could not distribute it between the number of workers in their own respected role.}'' 
Some others noted that they identified many factors that were necessary to consider when allocating resources but needed to simplify it to fit in their program. Highlighting this challenge of having to make a reductive decision, a student noted, ``\textit{I faced the challenge of programming something that would evaluate everyone's situations, and narrow it down to people getting priority over others.}'' 

\vspace{-0.5em}
\subsection{Learning and Conceptual Integration }

Throughout the semester, we made specific moves to link programming with social and relatable elements from students' daily lives. 
This integrated approach was well-received by many students who saw value in merging programming with social challenges as heard in a student's comment, ``\textit{It was meaningful to solve problems directly related to reality through programming.}'' 
Analyzing the students' reflections, it is evident that the integration enabled synergistic gain:  programming helped deepen their understanding of the social problems, while attending to complex social problems helped in deepening their knowledge of programming.

\subsubsection{Programming deepened social understanding}

Asking students to breakdown the problem such that a computing system can solve it created opportunities for the students to delve deeper into the social problem.
Reflecting on Project 2, a student remarked, ``\textit{I thought it was an exciting way of putting our programming knowledge into an actual world situation that many have experienced before. The ability to put it into code expanded my way of interpreting problems moving forward.}''
Computing, here, was perceived as a tool to understand the social problem as shared by another student, ``\textit{I think the main thing I learned was how programming is used to look into real ethical problems, which I think is really cool.}''

Responses varied on the challenges posed by the integrated problems. 
One student found the blend of social with programming novel and intriguing, ``\textit{I really liked the project, I think it tweaked with my brain and incorporating social issues with programming is something new to me.}''  
However, another student pointed out the inherent difficulty in translating intricate social issues into code, noting, ``\textit{...the social element incorporated by theory is valid but when embedded in the program it is so hard to quantify it.}''

\vspace{-0.5em}
\subsubsection{Social context helped deepen programming knowledge}

Embedding social challenges within programming exercises enhanced students' grasp of fundamental Java concepts. One student highlighted this improved understanding after Project 1, noting,  ``\textit{After doing the project I learned the `if' `else' in Java more.}'' 

We structured the programming tasks to emphasize the importance of planning \cite{loksa2016programming}: students first outlined their approach in English before translating it into Java. 
This often necessitated multiple iterations, where the students refined their solutions each time, going back and forth between the social and technical aspects of the assignment.
We believe that the iterative approach helps in deepening the students' programming abilities, a sentiment echoed by a student's remark on Project 3, ``\textit{I learned to reconcile ideas of justice with programming concepts. I became much more confident writing methods after coding the 4 `Moogle' filtering methods.}''

Many students shared the resilience and personal growth they experienced while working on the programming problems. One student reflected, ``\textit{I had some struggles but I managed to overcome them with time. ... spoke aloud on what my code was supposed to do which helped me identify the problem I encoded into my algorithm.}'' 

Overall, the students appreciated the socially responsible computing course.
   A student, looking back on the semester, commented, ``\textit{I like the class a lot and wasn't expecting learning so much about how technology and computing can affect people in society which was a nice surprise, I thought it would be more focused on programming which it usually is most of time except the first couple weeks and the last couple weeks but yet again I enjoy learning about the ethics and power computers have in out society.}''
Additionally, some students expressed a desire for a more personalized exploration of societal issues. One student suggested, ``\textit{I think it would be very fun to have a project that allows} [us] \textit{to pick an issue that interest us individually.}''

\vspace{-0.5em}
\section{Discussion and Conclusion}


Reflecting on our experience, we highlight some of the challenges that we faced. We acknowledge that these challenges are not exhaustive, nor do we have a generalizable solutions for them. 

\smallsection{Challenge 0: Building trust with and among students}
In examining social complexities and problems, our pedagogical approach leaned toward fostering a communal sense of inquiry. 
This was rooted in our recognition of the limitations of our knowledge about social problems and the importance of students exploring the principles that guide their thinking without necessarily reaching a convergent single answer. 
It is imperative, thus, to nurture trust both between the educator and students \emph{and} amongst the students themselves to encourage open dialogue.

Building trust is continuous, demanding consistent attention to the evolving classroom dynamics.
For this, we were taking cues from Learning for Justice, employing tactics such as urging students to contribute personal experiences, thereby enriching the discussions  \cite{learningforjustice}. 
Reflecting upon our sessions and revisiting our lectures, we recognize lapses in fostering inclusive conversations, particularly in empowering every student to voice their perspective.

\smallsection{Challenge 1: Being vulnerable to engage in the discussion}
Classroom discussions require students to trust each other \emph{and} the instructor so that they can remain open to divergent perspectives \cite{henry1994there}. 
Encouraging discourse with differing perspectives and ensuring that the students are engaging with perspectives different than theirs is quite challenging. 
``To teach is to be vulnerable'' \cite{bullough2005teacher}; we found ourselves noting that we needed to be vulnerable in the class, to say to the students that we did not know the answers and that there is no right answer. 
On a reflexive note, the instructor is a computer scientist with limited education in humanities and social sciences; the positivist training he had received throughout his formal education often gave rise to doubts in his ability to teach.
Many other CS instructors are in a similar position \cite{quinn2006teaching, johnson1994should}.
Reflecting on our experience, we believe that to succeed, we need to stay with the trouble \cite{haraway2016staying}, hold ourselves accountable \cite{washington2020twice}, and acknowledge the mistakes we make on a journey of constant learning \cite{evans2020mistakes}. 

In addition, we made moves to ensure that students were carefully evaluating their approaches and considering who their approach prioritized and ignored. 
We also ensured that they knew about our position, and clarified that we were sharing not to indoctrinate them with a singular belief but rather to guide them in thinking about their \emph{own} principles. 
For example, students had different views on what constitutes fairness and justice when deciding on-campus housing. During the discussion, the instructor needed to facilitate by steering students back to the principles that inform the differing notions of fairness, in this case, equality versus equity.

\smallsection{Challenge 2: Dovetailing technical problems with social issues}
By aligning programming challenges with real-world social problems, we aimed to bridge the gap between abstract technical concepts and their tangible impact on society, one that would be relatable and meaningful to the students. Our findings indicate that this synergy holds promising potential. 
Yet, weaving these domains introduces various ethical and practical complexities. A key concern is ensuring students can delve into and master the technical aspects within the provided timeframe while also understanding the nuances of societal issues. We attempted to strike a balance by incorporating group work and offering additional guidance, such as expected outputs. 
But the challenge remains.

Another challenge that we faced was figuring out how the students could demonstrate that they have attended to the social issue in sufficient depth through their programs, a challenge echoed in prior work (e.g., \cite{moskal2002grading, cohen2021new, kopec2023effectiveness, brodley2022broadening}). In the projects, we asked students to write reflections in two places, once before building their approach and once after they have developed it. Along with this, they wrote and explained test cases that would allow us to evaluate their approach. However, we struggled to evaluate some of the submissions, requiring prolonged deliberations between the graders and the instructor. 
Building rubrics akin to reflective essay evaluation would help. 
For future iterations, we will seek advice from instructors in liberal arts and the humanities who regularly develop such rubrics.

\smallsection{Challenge 3: Finding community stakeholder(s)}
Interactions beyond the classroom helped deepen students' understanding of the complexities surrounding societal issues. 
If we want our students to learn about the responsibilities that come with the power of developing computing systems, we need to foster space where students can connect with users and stakeholders impacted by these systems. 
However, we struggled with an ethical challenge to do so. On one hand, we wanted students to build systems that could impact community members.  On the other hand, we were worried that the engagement would not be reciprocal; the community members' time and effort may not be justifiably paid back. We, in academia, have often been exploitative and extractive \cite{stoecker2009unheard}. 
To navigate this, we encouraged students to liaise with friends and family, hoping that pre-existing bonds would provide clarity on the mutual benefit.

\smallsection{Challenge 4: Delineating the scope of technology-driven solutions}
One significant challenge we faced was ensuring students did not overstate the capabilities of their solutions. We needed to ensure that we scoped the problem such that the students could attend to it with their newly learned programming abilities. More critically, we needed to structure the assignments such that the larger problem at hand was not made reductive, allowing for complexities that go beyond technological solutions to emerge. For example, in Project 1, we asked students to come up with an approach that is fairer than the existing one, but at the end, we asked them to reflect on who their approach advantages and who is likely to be at a disadvantage. Such reflection not only allowed us to delineate the scope of the technology-driven solution but also allowed us to show the iterative process of building technological systems.

\smallsection{Challenge 5: Positive Societal Impact != Social Relevance}
Emphasizing social relevance requires attending to the contextual intricacies, power dynamics, and underlying systemic issues that  underpin societal challenges. 
While highlighting the positive impacts of computing can inspire students to engage actively in addressing real-world issues through technology, there is an inherent challenge for educators to not exclusively focus on these positive aspects. Adopting a balanced view is necessary \cite{ko2020time}. 
We need to incorporate critical analysis that provides a structured way to understand the broader societal impact of technology \cite{vakil2018ethics, medina2021towards, costanza2020design, eubanks2018automating, karetai2023decolonising}. 
Only then it will open learning opportunities for us to acknowledge the multifaceted nature of the relationship between computing and social change.  
Our findings further reveal that incorporating such critical approaches can help guide students toward a more comprehensive grasp of the societal impact of computing solutions, and to understand their role and responsibilities.

\smallsection{Challenge 6: Avoiding Responsibilization}
While the emphasis on individual responsibilities for ethical design is widespread, there is notable silence on corporate accountability.
Judith Butler terms this phenomenon ``responsibilization'' \cite{butler2016frames}. 
For a truly fair society, the focus must shift from responsibilization to making larger entities accountable, achieved through collective action and political change \cite{costanza2020design, karetai2023decolonising}.
This redirection from undue responsibilization is even more pressing 
given that many of our students are from lower-SES backgrounds
and tech jobs are often presented as the only pathway to economic mobility. 
In teaching socially responsible computing, a challenge lies in ensuring students grasp both the importance as well as limitations of individual responsibilities. 
In our course, we discussed individual responsibilities, the power developers have in refusing to build or in repairing harm, conflicts arising from such choices, and the power and responsibilities accorded to larger structures. 
We wish we had more opportunities to discuss these issues in detail.

In conclusion, our future computer scientists must be technically proficient \emph{and} socially and ethically conscious, becoming justice-seeking citizens. 
Our students' feedback, that the weaving of social and technical motivated and empowered them, is heartening and points towards a promising direction.
We hope that the insights shared in this article offer initial considerations for educators.

\vspace{-1em}
\begin{acks}
  This work is supported by NSF award 2216575 and a grant from Northeastern University's Center for Inclusive Computing. 
\end{acks}

\bibliographystyle{ACM-Reference-Format}
\bibliography{sample}


\begin{thebibliography}{51}


\ifx \showCODEN    \undefined \def \showCODEN     #1{\unskip}     \fi
\ifx \showDOI      \undefined \def \showDOI       #1{#1}\fi
\ifx \showISBNx    \undefined \def \showISBNx     #1{\unskip}     \fi
\ifx \showISBNxiii \undefined \def \showISBNxiii  #1{\unskip}     \fi
\ifx \showISSN     \undefined \def \showISSN      #1{\unskip}     \fi
\ifx \showLCCN     \undefined \def \showLCCN      #1{\unskip}     \fi
\ifx \shownote     \undefined \def \shownote      #1{#1}          \fi
\ifx \showarticletitle \undefined \def \showarticletitle #1{#1}   \fi
\ifx \showURL      \undefined \def \showURL       {\relax}        \fi
\providecommand\bibfield[2]{#2}
\providecommand\bibinfo[2]{#2}
\providecommand\natexlab[1]{#1}
\providecommand\showeprint[2][]{arXiv:#2}

\bibitem[Allen(2022)]%
        {allen2022feminist}
\bibfield{author}{\bibinfo{person}{Amy Allen}.}
  \bibinfo{year}{2022}\natexlab{}.
\newblock \showarticletitle{{Feminist Perspectives on Power}}.
\newblock In \bibinfo{booktitle}{\emph{The {Stanford} Encyclopedia of
  Philosophy} (\bibinfo{edition}{{F}all 2022} ed.)},
  \bibfield{editor}{\bibinfo{person}{Edward~N. Zalta} {and}
  \bibinfo{person}{Uri Nodelman}} (Eds.). \bibinfo{publisher}{Metaphysics
  Research Lab, Stanford University}.
\newblock


\bibitem[Boucher et~al\mbox{.}(2017)]%
        {boucher2017can}
\bibfield{author}{\bibinfo{person}{Kathryn~L Boucher},
  \bibinfo{person}{Melissa~A Fuesting}, \bibinfo{person}{Amanda~B Diekman},
  {and} \bibinfo{person}{Mary~C Murphy}.} \bibinfo{year}{2017}\natexlab{}.
\newblock \showarticletitle{Can I work with and help others in this field? How
  communal goals influence interest and participation in STEM fields}.
\newblock \bibinfo{journal}{\emph{Frontiers in psychology}}
  \bibinfo{volume}{8} (\bibinfo{year}{2017}), \bibinfo{pages}{901}.
\newblock


\bibitem[Brinkman and Diekman(2016)]%
        {brinkman2016applying}
\bibfield{author}{\bibinfo{person}{Bo Brinkman} {and} \bibinfo{person}{Amanda
  Diekman}.} \bibinfo{year}{2016}\natexlab{}.
\newblock \showarticletitle{Applying the communal goal congruity perspective to
  enhance diversity and inclusion in undergraduate computing degrees}. In
  \bibinfo{booktitle}{\emph{Proceedings of the 47th ACM technical symposium on
  computing science education}}. \bibinfo{pages}{102--107}.
\newblock


\bibitem[Brodley et~al\mbox{.}(2022)]%
        {brodley2022broadening}
\bibfield{author}{\bibinfo{person}{Carla~E Brodley},
  \bibinfo{person}{Benjamin~J Hescott}, \bibinfo{person}{Jessica Biron},
  \bibinfo{person}{Ali Ressing}, \bibinfo{person}{Melissa Peiken},
  \bibinfo{person}{Sarah Maravetz}, {and} \bibinfo{person}{Alan Mislove}.}
  \bibinfo{year}{2022}\natexlab{}.
\newblock \showarticletitle{Broadening participation in computing via
  ubiquitous combined majors (CS+ X)}. In \bibinfo{booktitle}{\emph{Proceedings
  of the 53rd ACM Technical Symposium on Computer Science Education-Volume 1}}.
  \bibinfo{pages}{544--550}.
\newblock


\bibitem[Brown et~al\mbox{.}(2022)]%
        {brown2022shortest}
\bibfield{author}{\bibinfo{person}{Noelle Brown}, \bibinfo{person}{Koriann
  South}, {and} \bibinfo{person}{Eliane~S Wiese}.}
  \bibinfo{year}{2022}\natexlab{}.
\newblock \showarticletitle{The Shortest Path to Ethics in AI: An Integrated
  Assignment Where Human Concerns Guide Technical Decisions}. In
  \bibinfo{booktitle}{\emph{Proceedings of the 2022 ACM Conference on
  International Computing Education Research-Volume 1}}.
  \bibinfo{pages}{344--355}.
\newblock


\bibitem[Buckley et~al\mbox{.}(2004)]%
        {buckley2004benefits}
\bibfield{author}{\bibinfo{person}{Michael Buckley}, \bibinfo{person}{Helene
  Kershner}, \bibinfo{person}{Kris Schindler}, \bibinfo{person}{Carl Alphonce},
  {and} \bibinfo{person}{Jennifer Braswell}.} \bibinfo{year}{2004}\natexlab{}.
\newblock \showarticletitle{Benefits of using socially-relevant projects in
  computer science and engineering education}. In
  \bibinfo{booktitle}{\emph{Proceedings of the 35th SIGCSE technical symposium
  on Computer science education}}. \bibinfo{pages}{482--486}.
\newblock


\bibitem[Bullough~Jr(2005)]%
        {bullough2005teacher}
\bibfield{author}{\bibinfo{person}{Robert~V Bullough~Jr}.}
  \bibinfo{year}{2005}\natexlab{}.
\newblock \showarticletitle{Teacher vulnerability and teachability: A case
  study of a mentor and two interns}.
\newblock \bibinfo{journal}{\emph{Teacher Education Quarterly}}
  (\bibinfo{year}{2005}), \bibinfo{pages}{23--39}.
\newblock


\bibitem[Butler(2016)]%
        {butler2016frames}
\bibfield{author}{\bibinfo{person}{Judith Butler}.}
  \bibinfo{year}{2016}\natexlab{}.
\newblock \bibinfo{booktitle}{\emph{Frames of war: When is life grievable?}}
\newblock \bibinfo{publisher}{Verso Books}.
\newblock


\bibitem[Califf and Goodwin(2005)]%
        {califf2005effective}
\bibfield{author}{\bibinfo{person}{Mary~Elaine Califf} {and}
  \bibinfo{person}{Mary Goodwin}.} \bibinfo{year}{2005}\natexlab{}.
\newblock \showarticletitle{Effective incorporation of ethics into courses that
  focus on programming}.
\newblock \bibinfo{journal}{\emph{ACM SIGCSE Bulletin}} \bibinfo{volume}{37},
  \bibinfo{number}{1} (\bibinfo{year}{2005}).
\newblock


\bibitem[Castro et~al\mbox{.}(2023)]%
        {castro2023piloting}
\bibfield{author}{\bibinfo{person}{Francisco Castro}, \bibinfo{person}{Sahitya
  Raipura}, \bibinfo{person}{Heather Conboy}, \bibinfo{person}{Peter Haas},
  \bibinfo{person}{Leon Osterweil}, {and} \bibinfo{person}{Ivon Arroyo}.}
  \bibinfo{year}{2023}\natexlab{}.
\newblock \showarticletitle{Piloting an Interactive Ethics and Responsible
  Computing Learning Environment in Undergraduate CS Courses}. In
  \bibinfo{booktitle}{\emph{Proceedings of the 54th ACM Technical Symposium on
  Computer Science Education V. 1}}. \bibinfo{pages}{659--665}.
\newblock


\bibitem[Cohen et~al\mbox{.}(2021)]%
        {cohen2021new}
\bibfield{author}{\bibinfo{person}{Lena Cohen}, \bibinfo{person}{Heila Precel},
  \bibinfo{person}{Harold Triedman}, {and} \bibinfo{person}{Kathi Fisler}.}
  \bibinfo{year}{2021}\natexlab{}.
\newblock \showarticletitle{A new model for weaving responsible computing into
  courses across the CS curriculum}. In \bibinfo{booktitle}{\emph{Proceedings
  of the 52nd ACM Technical Symposium on Computer Science Education}}.
\newblock


\bibitem[Connolly(2011)]%
        {connolly2011beyond}
\bibfield{author}{\bibinfo{person}{Randy~W Connolly}.}
  \bibinfo{year}{2011}\natexlab{}.
\newblock \showarticletitle{Beyond good and evil impacts: Rethinking the social
  issues components in our computing curricula}. In
  \bibinfo{booktitle}{\emph{Proceedings of the 16th annual joint conference on
  Innovation and technology in computer science education}}.
\newblock


\bibitem[Corbett and Hill(2015)]%
        {corbett2015solving}
\bibfield{author}{\bibinfo{person}{Christianne Corbett} {and}
  \bibinfo{person}{Catherine Hill}.} \bibinfo{year}{2015}\natexlab{}.
\newblock \bibinfo{booktitle}{\emph{Solving the Equation: The Variables for
  Women's Success in Engineering and Computing.}}
\newblock \bibinfo{publisher}{ERIC}.
\newblock


\bibitem[Costanza-Chock(2020)]%
        {costanza2020design}
\bibfield{author}{\bibinfo{person}{Sasha Costanza-Chock}.}
  \bibinfo{year}{2020}\natexlab{}.
\newblock \bibinfo{booktitle}{\emph{Design justice: Community-led practices to
  build the worlds we need}}.
\newblock \bibinfo{publisher}{The MIT Press}.
\newblock


\bibitem[Council et~al\mbox{.}(2013)]%
        {national2013nonresponse}
\bibfield{author}{\bibinfo{person}{National~Research Council} {et~al\mbox{.}}}
  \bibinfo{year}{2013}\natexlab{}.
\newblock \showarticletitle{Nonresponse in social science surveys: A research
  agenda}.
\newblock  (\bibinfo{year}{2013}).
\newblock


\bibitem[Dean and Nourbakhsh(2022)]%
        {dean2022teaching}
\bibfield{author}{\bibinfo{person}{Victoria Dean} {and} \bibinfo{person}{Illah
  Nourbakhsh}.} \bibinfo{year}{2022}\natexlab{}.
\newblock \showarticletitle{Teaching Ethics by Teaching Ethics Pedagogy: A
  Proposal for Structural Ethics Intervention}. In
  \bibinfo{booktitle}{\emph{Proceedings of the 53rd ACM Technical Symposium on
  Computer Science Education-Volume 1}}. \bibinfo{pages}{272--278}.
\newblock


\bibitem[Diekman et~al\mbox{.}(2010)]%
        {diekman2010seeking}
\bibfield{author}{\bibinfo{person}{Amanda~B Diekman},
  \bibinfo{person}{Elizabeth~R Brown}, \bibinfo{person}{Amanda~M Johnston},
  {and} \bibinfo{person}{Emily~K Clark}.} \bibinfo{year}{2010}\natexlab{}.
\newblock \showarticletitle{Seeking congruity between goals and roles: A new
  look at why women opt out of science, technology, engineering, and
  mathematics careers}.
\newblock \bibinfo{journal}{\emph{Psychological science}} \bibinfo{volume}{21},
  \bibinfo{number}{8} (\bibinfo{year}{2010}), \bibinfo{pages}{1051--1057}.
\newblock


\bibitem[Eubanks(2018)]%
        {eubanks2018automating}
\bibfield{author}{\bibinfo{person}{Virginia Eubanks}.}
  \bibinfo{year}{2018}\natexlab{}.
\newblock \bibinfo{booktitle}{\emph{Automating inequality: How high-tech tools
  profile, police, and punish the poor}}.
\newblock \bibinfo{publisher}{St. Martin's Press}.
\newblock


\bibitem[Evans-Santiago(2020)]%
        {evans2020mistakes}
\bibfield{author}{\bibinfo{person}{Bre Evans-Santiago}.}
  \bibinfo{year}{2020}\natexlab{}.
\newblock \bibinfo{booktitle}{\emph{Mistakes we have made: Implications for
  social justice educators}}.
\newblock \bibinfo{publisher}{Myers Education Press}.
\newblock


\bibitem[Fiesler et~al\mbox{.}(2021)]%
        {fiesler2021integrating}
\bibfield{author}{\bibinfo{person}{Casey Fiesler}, \bibinfo{person}{Mikhaila
  Friske}, \bibinfo{person}{Natalie Garrett}, \bibinfo{person}{Felix Muzny},
  \bibinfo{person}{Jessie~J Smith}, {and} \bibinfo{person}{Jason Zietz}.}
  \bibinfo{year}{2021}\natexlab{}.
\newblock \showarticletitle{Integrating ethics into introductory programming
  classes}. In \bibinfo{booktitle}{\emph{Proceedings of the 52nd ACM Technical
  Symposium on Computer Science Education}}.
\newblock


\bibitem[Fiesler et~al\mbox{.}(2020)]%
        {fiesler2020we}
\bibfield{author}{\bibinfo{person}{Casey Fiesler}, \bibinfo{person}{Natalie
  Garrett}, {and} \bibinfo{person}{Nathan Beard}.}
  \bibinfo{year}{2020}\natexlab{}.
\newblock \showarticletitle{What do we teach when we teach tech ethics? A
  syllabi analysis}. In \bibinfo{booktitle}{\emph{Proceedings of the 51st ACM
  technical symposium on computer science education}}.
  \bibinfo{pages}{289--295}.
\newblock


\bibitem[for Justice({[n.\,d.]})]%
        {learningforjustice}
\bibfield{author}{\bibinfo{person}{Learning for Justice}.}
  \bibinfo{year}{[n.\,d.]}\natexlab{}.
\newblock \bibinfo{title}{Community Inquiry}.
\newblock
  \bibinfo{howpublished}{\url{https://www.learningforjustice.org/classroom-resources/teaching-strategies/community-inquiry}}.
\newblock
\newblock
\shownote{Accessed: July 26, 2023}.


\bibitem[Goldweber et~al\mbox{.}(2011)]%
        {goldweber2011enhancing}
\bibfield{author}{\bibinfo{person}{Mikey Goldweber}, \bibinfo{person}{Renzo
  Davoli}, \bibinfo{person}{Joyce~Currie Little}, \bibinfo{person}{Charles
  Riedesel}, \bibinfo{person}{Henry Walker}, \bibinfo{person}{Gerry Cross},
  {and} \bibinfo{person}{Brian~R Von~Konsky}.} \bibinfo{year}{2011}\natexlab{}.
\newblock \showarticletitle{Enhancing the social issues components in our
  computing curriculum: computing for the social good}.
\newblock \bibinfo{journal}{\emph{ACM Inroads}} \bibinfo{volume}{2},
  \bibinfo{number}{1} (\bibinfo{year}{2011}), \bibinfo{pages}{64--82}.
\newblock


\bibitem[Haraway(2016)]%
        {haraway2016staying}
\bibfield{author}{\bibinfo{person}{Donna~J Haraway}.}
  \bibinfo{year}{2016}\natexlab{}.
\newblock \bibinfo{booktitle}{\emph{Staying with the trouble: Making kin in the
  Chthulucene}}.
\newblock \bibinfo{publisher}{Duke University Press}.
\newblock


\bibitem[Henry(1994)]%
        {henry1994there}
\bibfield{author}{\bibinfo{person}{Annette Henry}.}
  \bibinfo{year}{1994}\natexlab{}.
\newblock \showarticletitle{There are no safe places: Pedagogy as powerful and
  dangerous terrain}.
\newblock \bibinfo{journal}{\emph{Action in Teacher Education}}
  \bibinfo{volume}{15}, \bibinfo{number}{4} (\bibinfo{year}{1994}),
  \bibinfo{pages}{1--4}.
\newblock


\bibitem[Horton et~al\mbox{.}(2023)]%
        {horton2023more}
\bibfield{author}{\bibinfo{person}{Diane Horton}, \bibinfo{person}{David Liu},
  \bibinfo{person}{Sheila~A McIlraith}, {and} \bibinfo{person}{Nina Wang}.}
  \bibinfo{year}{2023}\natexlab{}.
\newblock \showarticletitle{Is More Better When Embedding Ethics in CS
  Courses?}. In \bibinfo{booktitle}{\emph{Proceedings of the 54th ACM Technical
  Symposium on Computer Science Education V. 1}}. \bibinfo{pages}{652--658}.
\newblock


\bibitem[Johnson(1994)]%
        {johnson1994should}
\bibfield{author}{\bibinfo{person}{Deborah Johnson}.}
  \bibinfo{year}{1994}\natexlab{}.
\newblock \showarticletitle{Who should teach computer ethics and computers \&
  society?}
\newblock \bibinfo{journal}{\emph{Acm Sigcas Computers and Society}}
  \bibinfo{volume}{24}, \bibinfo{number}{2} (\bibinfo{year}{1994}),
  \bibinfo{pages}{6--13}.
\newblock


\bibitem[Karetai et~al\mbox{.}(2023)]%
        {karetai2023decolonising}
\bibfield{author}{\bibinfo{person}{Mawera Karetai}, \bibinfo{person}{Samuel
  Mann}, \bibinfo{person}{Dhammika~Dave Guruge}, \bibinfo{person}{Sherlock
  Licorish}, {and} \bibinfo{person}{Alison Clear}.}
  \bibinfo{year}{2023}\natexlab{}.
\newblock \showarticletitle{Decolonising Computer Science Education-A Global
  Perspective}. In \bibinfo{booktitle}{\emph{Proceedings of the 54th ACM
  Technical Symposium on Computer Science Education V.1}}.
  \bibinfo{pages}{1097--1102}.
\newblock


\bibitem[Kiesler and Thorbr{\"u}gge(2023)]%
        {kiesler2023socially}
\bibfield{author}{\bibinfo{person}{Natalie Kiesler} {and}
  \bibinfo{person}{Carsten Thorbr{\"u}gge}.} \bibinfo{year}{2023}\natexlab{}.
\newblock \showarticletitle{Socially responsible programming in computing
  education and expectations in the profession}. In
  \bibinfo{booktitle}{\emph{Proceedings of the 2023 Conference on Innovation
  and Technology in Computer Science Education V.1}}.
\newblock


\bibitem[Ko et~al\mbox{.}(2020)]%
        {ko2020time}
\bibfield{author}{\bibinfo{person}{Amy~J Ko}, \bibinfo{person}{Alannah Oleson},
  \bibinfo{person}{Neil Ryan}, \bibinfo{person}{Yim Register},
  \bibinfo{person}{Benjamin Xie}, \bibinfo{person}{Mina Tari},
  \bibinfo{person}{Matthew Davidson}, \bibinfo{person}{Stefania Druga}, {and}
  \bibinfo{person}{Dastyni Loksa}.} \bibinfo{year}{2020}\natexlab{}.
\newblock \showarticletitle{It is time for more critical CS education}.
\newblock \bibinfo{journal}{\emph{Commun. ACM}} \bibinfo{volume}{63},
  \bibinfo{number}{11} (\bibinfo{year}{2020}), \bibinfo{pages}{31--33}.
\newblock


\bibitem[Kopec et~al\mbox{.}(2023)]%
        {kopec2023effectiveness}
\bibfield{author}{\bibinfo{person}{Matthew Kopec}, \bibinfo{person}{Meica
  Magnani}, \bibinfo{person}{Vance Ricks}, \bibinfo{person}{Roben Torosyan},
  \bibinfo{person}{John Basl}, \bibinfo{person}{Nicholas Miklaucic},
  \bibinfo{person}{Felix Muzny}, \bibinfo{person}{Ronald Sandler},
  \bibinfo{person}{Christo Wilson}, \bibinfo{person}{Adam Wisniewski-Jensen},
  {et~al\mbox{.}}} \bibinfo{year}{2023}\natexlab{}.
\newblock \showarticletitle{The effectiveness of embedded values analysis
  modules in Computer Science education: An empirical study}.
\newblock \bibinfo{journal}{\emph{Big Data \& Society}} \bibinfo{volume}{10},
  \bibinfo{number}{1} (\bibinfo{year}{2023}),
  \bibinfo{pages}{20539517231176230}.
\newblock


\bibitem[Loksa et~al\mbox{.}(2016)]%
        {loksa2016programming}
\bibfield{author}{\bibinfo{person}{Dastyni Loksa}, \bibinfo{person}{Amy~J Ko},
  \bibinfo{person}{Will Jernigan}, \bibinfo{person}{Alannah Oleson},
  \bibinfo{person}{Christopher~J Mendez}, {and} \bibinfo{person}{Margaret~M
  Burnett}.} \bibinfo{year}{2016}\natexlab{}.
\newblock \showarticletitle{Programming, problem solving, and self-awareness:
  Effects of explicit guidance}. In \bibinfo{booktitle}{\emph{Proceedings of
  the 2016 CHI conference on human factors in computing systems}}.
  \bibinfo{pages}{1449--1461}.
\newblock


\bibitem[Martin and Weltz(1999)]%
        {martin1999awareness}
\bibfield{author}{\bibinfo{person}{C~Dianne Martin} {and}
  \bibinfo{person}{Elaine~Yale Weltz}.} \bibinfo{year}{1999}\natexlab{}.
\newblock \showarticletitle{From awareness to action: Integrating ethics and
  social responsibility into the computer science curriculum}.
\newblock \bibinfo{journal}{\emph{ACM Sigcas Computers and Society}}
  \bibinfo{volume}{29}, \bibinfo{number}{2} (\bibinfo{year}{1999}),
  \bibinfo{pages}{6--14}.
\newblock


\bibitem[Medina-Kim(2021)]%
        {medina2021towards}
\bibfield{author}{\bibinfo{person}{Gabriel Medina-Kim}.}
  \bibinfo{year}{2021}\natexlab{}.
\newblock \showarticletitle{Towards Justice in Undergraduate Computer Science
  Education: Possibilities in Power, Equity, and Praxis}. In
  \bibinfo{booktitle}{\emph{2021 ASEE Virtual Annual Conference Content
  Access}}.
\newblock


\bibitem[Moskal et~al\mbox{.}(2002)]%
        {moskal2002grading}
\bibfield{author}{\bibinfo{person}{Barbara Moskal}, \bibinfo{person}{Keith
  Miller}, {and} \bibinfo{person}{LA~Smith King}.}
  \bibinfo{year}{2002}\natexlab{}.
\newblock \showarticletitle{Grading essays in computer ethics: rubrics
  considered helpful}. In \bibinfo{booktitle}{\emph{Proceedings of the 33rd
  SIGCSE technical symposium on Computer science education}}.
  \bibinfo{pages}{101--105}.
\newblock


\bibitem[Pauca and Guy(2012)]%
        {pauca2012mobile}
\bibfield{author}{\bibinfo{person}{Victor~Paul Pauca} {and}
  \bibinfo{person}{Richard~T Guy}.} \bibinfo{year}{2012}\natexlab{}.
\newblock \showarticletitle{Mobile apps for the greater good: a socially
  relevant approach to software engineering}. In
  \bibinfo{booktitle}{\emph{Proceedings of the 43rd ACM technical symposium on
  Computer Science Education}}. \bibinfo{pages}{535--540}.
\newblock


\bibitem[Peck(2017)]%
        {peck2017ethical}
\bibfield{author}{\bibinfo{person}{Evan Peck}.}
  \bibinfo{year}{2017}\natexlab{}.
\newblock \showarticletitle{The Ethical Engine: Integrating Ethical Design into
  Intro Computer Science}.
\newblock \bibinfo{journal}{\emph{blog, Bucknell HCI}}  \bibinfo{volume}{5}
  (\bibinfo{year}{2017}).
\newblock


\bibitem[Pinkard et~al\mbox{.}(2020)]%
        {pinkard2020equitable}
\bibfield{author}{\bibinfo{person}{Nichole Pinkard},
  \bibinfo{person}{Caitlin~Kennedy Martin}, {and} \bibinfo{person}{Sheena
  Erete}.} \bibinfo{year}{2020}\natexlab{}.
\newblock \showarticletitle{Equitable approaches: Opportunities for
  computational thinking with emphasis on creative production and connections
  to community}.
\newblock \bibinfo{journal}{\emph{Interactive Learning Environments}}
  \bibinfo{volume}{28}, \bibinfo{number}{3} (\bibinfo{year}{2020}),
  \bibinfo{pages}{347--361}.
\newblock


\bibitem[Quinn(2006)]%
        {quinn2006teaching}
\bibfield{author}{\bibinfo{person}{Michael~J Quinn}.}
  \bibinfo{year}{2006}\natexlab{}.
\newblock \showarticletitle{On teaching computer ethics within a computer
  science department}.
\newblock \bibinfo{journal}{\emph{Science and Engineering Ethics}}
  \bibinfo{volume}{12} (\bibinfo{year}{2006}), \bibinfo{pages}{335--343}.
\newblock


\bibitem[Ryoo et~al\mbox{.}(2021)]%
        {ryoo2021happens}
\bibfield{author}{\bibinfo{person}{Jean~J Ryoo}, \bibinfo{person}{Alicia
  Morris}, {and} \bibinfo{person}{Jane Margolis}.}
  \bibinfo{year}{2021}\natexlab{}.
\newblock \showarticletitle{“What happens to the Raspado man in a cash-free
  society?”: Teaching and learning socially responsible computing}.
\newblock \bibinfo{journal}{\emph{ACM Transactions on Computing Education
  (TOCE)}} \bibinfo{volume}{21}, \bibinfo{number}{4} (\bibinfo{year}{2021}),
  \bibinfo{pages}{1--28}.
\newblock


\bibitem[Salda{\~n}a(2015)]%
        {saldana2015coding}
\bibfield{author}{\bibinfo{person}{Johnny Salda{\~n}a}.}
  \bibinfo{year}{2015}\natexlab{}.
\newblock \bibinfo{booktitle}{\emph{The coding manual for qualitative
  researchers}}.
\newblock \bibinfo{publisher}{Sage}.
\newblock


\bibitem[Saltz et~al\mbox{.}(2019)]%
        {saltz2019integrating}
\bibfield{author}{\bibinfo{person}{Jeffrey Saltz}, \bibinfo{person}{Michael
  Skirpan}, \bibinfo{person}{Casey Fiesler}, \bibinfo{person}{Micha Gorelick},
  \bibinfo{person}{Tom Yeh}, \bibinfo{person}{Robert Heckman},
  \bibinfo{person}{Neil Dewar}, {and} \bibinfo{person}{Nathan Beard}.}
  \bibinfo{year}{2019}\natexlab{}.
\newblock \showarticletitle{Integrating ethics within machine learning
  courses}.
\newblock \bibinfo{journal}{\emph{ACM Transactions on Computing Education
  (TOCE)}} \bibinfo{volume}{19}, \bibinfo{number}{4} (\bibinfo{year}{2019}).
\newblock


\bibitem[Smith et~al\mbox{.}(2023)]%
        {smith2023incorporating}
\bibfield{author}{\bibinfo{person}{Jessie~J Smith}, \bibinfo{person}{Blakeley~H
  Payne}, \bibinfo{person}{Shamika Klassen}, \bibinfo{person}{Dylan~Thomas
  Doyle}, {and} \bibinfo{person}{Casey Fiesler}.}
  \bibinfo{year}{2023}\natexlab{}.
\newblock \showarticletitle{Incorporating Ethics in Computing Courses:
  Barriers, Support, and Perspectives from Educators}. In
  \bibinfo{booktitle}{\emph{Proceedings of the 54th ACM Technical Symposium on
  Computer Science Education V. 1}}. \bibinfo{pages}{367--373}.
\newblock


\bibitem[Stewart et~al\mbox{.}(2022)]%
        {stewart2022equity}
\bibfield{author}{\bibinfo{person}{Kylan Stewart}, \bibinfo{person}{Bruce
  Debruhl}, {and} \bibinfo{person}{Zoe Wood}.} \bibinfo{year}{2022}\natexlab{}.
\newblock \showarticletitle{An Equity-minded Assessment of Belonging among
  Computing Students}. In \bibinfo{booktitle}{\emph{2022 ASEE Annual Conference
  \& Exposition}}.
\newblock


\bibitem[Stoecker and Tryon(2009)]%
        {stoecker2009unheard}
\bibfield{author}{\bibinfo{person}{Randy Stoecker} {and}
  \bibinfo{person}{Elizabeth~A Tryon}.} \bibinfo{year}{2009}\natexlab{}.
\newblock \bibinfo{booktitle}{\emph{The unheard voices: Community organizations
  and service learning}}.
\newblock \bibinfo{publisher}{Temple University Press}.
\newblock


\bibitem[Tucker et~al\mbox{.}(1991)]%
        {curricula1991report}
\bibfield{author}{\bibinfo{person}{Allen~B. Tucker}, \bibinfo{person}{Bruce~H.
  Barnes}, \bibinfo{person}{Robert~M. Aiken}, \bibinfo{person}{Keith Barker},
  \bibinfo{person}{J. Bruce, Kim B. Thomas~Cain}, \bibinfo{person}{Susan~E.
  Conry}, \bibinfo{person}{Gerald~L. Engel}, \bibinfo{person}{Richard~G.
  Epstein}, \bibinfo{person}{Doris~K. Lidtke}, \bibinfo{person}{Michael~C.
  Mulder}, \bibinfo{person}{Jean~B. Rogers}, \bibinfo{person}{Eugene~H.
  Spafford}, {and} \bibinfo{person}{A. Joe~Turner}.}
  \bibinfo{year}{1991}\natexlab{}.
\newblock \showarticletitle{Report of the ACM/IEEE-CS Joint Curriculum Task
  Force}.
\newblock \bibinfo{journal}{\emph{Association for Computing Machinery}}
  (\bibinfo{year}{1991}).
\newblock
\showISBNx{089793817}


\bibitem[Vakil(2018)]%
        {vakil2018ethics}
\bibfield{author}{\bibinfo{person}{Sepehr Vakil}.}
  \bibinfo{year}{2018}\natexlab{}.
\newblock \showarticletitle{Ethics, identity, and political vision: Toward a
  justice-centered approach to equity in computer science education}.
\newblock \bibinfo{journal}{\emph{Harvard educational review}}
  \bibinfo{volume}{88}, \bibinfo{number}{1} (\bibinfo{year}{2018}),
  \bibinfo{pages}{26--52}.
\newblock


\bibitem[Washington(2020)]%
        {washington2020twice}
\bibfield{author}{\bibinfo{person}{Alicia~Nicki Washington}.}
  \bibinfo{year}{2020}\natexlab{}.
\newblock \showarticletitle{When twice as good isn't enough: The case for
  cultural competence in computing}. In \bibinfo{booktitle}{\emph{Proceedings
  of the 51st ACM technical symposium on computer science education}}.
  \bibinfo{pages}{213--219}.
\newblock


\bibitem[Werth(1997)]%
        {werth1997getting}
\bibfield{author}{\bibinfo{person}{Laurie~H Werth}.}
  \bibinfo{year}{1997}\natexlab{}.
\newblock \showarticletitle{Getting started with computer ethics}. In
  \bibinfo{booktitle}{\emph{Proceedings of the Twenty-eighth SIGCSE technical
  symposium on Computer Science Education}}. \bibinfo{pages}{1--5}.
\newblock


\bibitem[Yadav and Heath(2022)]%
        {yadav2022breaking}
\bibfield{author}{\bibinfo{person}{Aman Yadav} {and} \bibinfo{person}{Marie~K
  Heath}.} \bibinfo{year}{2022}\natexlab{}.
\newblock \showarticletitle{Breaking the code: Confronting racism in computer
  science through community, criticality, and citizenship}.
\newblock \bibinfo{journal}{\emph{TechTrends}} \bibinfo{volume}{66},
  \bibinfo{number}{3} (\bibinfo{year}{2022}), \bibinfo{pages}{450--458}.
\newblock


\bibitem[Zegura et~al\mbox{.}(2023)]%
        {zegura2023moment}
\bibfield{author}{\bibinfo{person}{Cass Zegura}, \bibinfo{person}{Ben~Rydal
  Shapiro}, \bibinfo{person}{Robert MacDonald}, \bibinfo{person}{Jason
  Borenstein}, {and} \bibinfo{person}{Ellen Zegura}.}
  \bibinfo{year}{2023}\natexlab{}.
\newblock \showarticletitle{“Moment to Moment”: A Situated View of Teaching
  Ethics from the Perspective of Computing Ethics Teaching Assistants}. In
  \bibinfo{booktitle}{\emph{Proceedings of the 2023 CHI Conference on Human
  Factors in Computing Systems}}. \bibinfo{pages}{1--15}.
\newblock


\end{thebibliography}


\end{document}